\documentclass[pre,normal]{revtex4}
\usepackage[dvips]{graphics}
\usepackage{epsfig}
\usepackage{epsfig}
\usepackage{dsfont}
\usepackage{amsmath}

\newcommand{\g}{\gamma}

\newcommand{\lgl}{\langle}
\newcommand{\rgl}{\rangle}

\begin{document}

\title{Towards correlated random networks}
\author{W. \surname{Pietsch}\footnote{wpietsch@gmx.de}}

\affiliation{Department for Philosophy of Science, University of Augsburg,
Universit\"atsstr.\ 10, D-86135 Augsburg, Germany}

\begin{abstract}
A model of correlated random networks is examined, i.e.\ networks with correlations between the degrees of neighboring nodes. These nodes do not necessarily have to be direct neighbors, the maximum range of the correlations can be arbitrarily chosen. Two different methods for the creation of such networks are presented: one of them is a generalization of a well-known algorithm by Maslov and Sneppen. The percolation threshold for the model is calculated and the result is tested using analytically solvable examples and simulations. In the end the principal importance of correlations and clustering for the topology of networks is discussed. Using a straight-forward extension of the network model by Barab\'asi and Albert, it is shown how a clustering-coefficient independent of the network size can originate in growing networks.\end{abstract}

\maketitle

\section{Introduction}

Scale-free networks, i.e.\ networks with essentially power-law degree distributions, have recently been widely studied (see \cite{baretalb} and \cite{dorogov} for reviews). Such degree distributions have been found in many different contexts, for example in several technological webs like the Internet \cite{int}, the WWW \cite{WWW}, or electrical power grids, and also in social networks, like the network of sexual contacts \cite{sex} or one of the phone calls \cite{phonecall}.

The standard model reproducing scale free degree distributions was introduced by Barab\'asi and Albert (BA-model) \cite{BA-model}. It is based on a growth algorithm with preferential attachment. A second older model which is also widely studied in the context of scale-free networks is the configuration model treated by Molloy and Reed (MR-model) \cite{mollreed}. This is to some extent the `most random' model possessing a given degree distribution $P(k)$ and a given number $N$ of nodes. The building prescription starts with sets of $NP(k)$ nodes with $k$ stubs each. The stubs are then connected randomly to each other; two connected stubs form a link. Double bonds and autoconnections can be neglected in the limit of large networks $N \rightarrow \infty$. MR-networks exhibit an arbitrary degree distribution but no correlations, i.e. the distribution of nodes on one end of a link is independent from that on the other end of the link. For the BA-model there are non-trivial correlations between neighboring nodes. For both networks the correlations cannot be influenced by the construction algorithm. In contrast to that, real networks exhibit a wide range of different correlations. For example, some are assortative \cite{Newm,49} while others are dissortative \cite{44,56} (i.e.\ nodes are generally attached to nodes with similar degree or not). It has been noted for some time, that the failure to include into the standard models this wide range of different correlations is a considerable defect of the models. So, a lot of effort has been put into the solution of this problem (e.g. \cite{11,12,13}).

This publication aims to contribute to this effort. We introduce a general model for correlations, the most natural generalization of the MR-model. To our knowledge this model has not yet been proposed in the full generality as we present it here. On the other hand, it has been examined in considerable detail for correlations between neighboring nodes. The generalization from correlations between neigboring nodes to those of arbitrary range might be considered only a small advance. However, since there is no full knowledge about the actual correlations in natural networks, correlations of long range should not be neglected a priori. As simple examples can prove, sometimes only correlations of long range have a decisive influence on important topological properties like the percolation threshold.

Thus far, an analytical expression for the percolation threshold has been derived only for a small class of random networks. In \cite{cohetal,mollreed} MR-networks are treated. In the present manuscript, we derive a general result for random networks with arbitrary correlations, i.e.\ correlations between direct neighbors but also between second neighbors etc. Employing the set of Eq.s (\ref{final1}) and (\ref{final2}) the percolation threshold can be calculated at least in principle. Related work can be found in \cite{Newm}. There, Newman studies the influence of mixing on the size of the giant component by mimicking percolation with the change of a characteristic degree scale. In \cite{Vaz} Vazquez and Moreno calculate the percolation threshold for random networks with correlations between direct neighbors. In a completely different way, we will also derive this threshold in Sec.~\ref{perer}. Contrary to our method Vazquez and Moreno can calculate the size of the giant component. Finally, in \cite{neu} Bogu\~n\'a and Serrano present a general theory for percolation in directed random networks with general two point correlations and bidirectional links.

\section{The model}
\label{modmod}
We distinguish different classes of networks by the range $l-1$ of the correlations: $P(k_{l+1}|k_l;...;k_1)$ depends on $k_2$, but not on $k_1$. The distribution $P(k_{l+1}|k_l;...;k_1)$ denotes the probability that a randomly chosen path which has the node sequence with degrees $k_1$, $k_2$, ..., $k_l$ is \emph{continued} by a node with degree $k_{l+1}$. The normalization condition is: $\sum_{k_{x}}P(k_{x}|k_{x-1};...;k_1)=1$. By a randomly chosen path we understand: `The first link of the path is chosen uniformly at random. The following link is chosen at random under the condition that it is adjacent to the first etc.' Consequently, the probability that a randomly chosen path of length $l-1$ has the node sequence $k_1$, $k_2$, ..., $k_{l}$ is
\begin{equation}
T(k_l;...;k_1)=  \frac{P(k_1) k_1 \left [\prod\limits_{s=2}^{l-1}   
P(k_s|k_{s-1};...;k_1)(k_s-1) \right ] P(k_{l}|k_{l-1};...;k_1)}
{ \sum\limits_{k_1,...,k_{l-1}} P(k_1) k_1 \left [ \prod\limits_{s=2}^{l-1}   
P(k_s|k_{s-1};...;k_1)(k_s-1) \right ]}.
\label{eq:tp}
\end{equation}
The factor $k_1 P(k_1)$ is proportional to the probability for the node-degree $k_1$ at the beginning, i.e. the first position of the path. $P(k_s|k_{s-1};...;k_1)(k_s-1)$ then denotes the probability that at position $s<l$ in the path we find a node with degree $k_s$. The last term $P(k_l|k_{l-1};...;k_1)$ accounts for the last position in the path. This time, the normalization is: $\sum_{k_l,...k_1}T(k_l;...;k_1)=1$. Of course, in all networks we need to have
\begin{equation} 
T(k_l;k_{l-1};...;k_2;k_1)=T(k_1;k_2;...;k_{l-1};k_l), \qquad \forall l \in \mathds{N}. \label{paths}
\end{equation}
Assuming a maximum range $l-1$ of the correlations in the network, only the paths of length $\leq l-1$ need to be considered. 

\paragraph*{An example}
\begin{figure}[tb]
\centering
\includegraphics[width=0.2\textwidth]{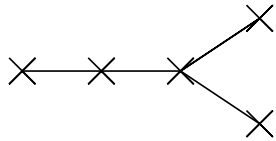}
\caption{An exemplary network.}
\label{fig:exam}
\end{figure}
In Fig.\ \ref{fig:exam} we see an exemplary network, in which we can measure the probabilities $T$. We can then determine recursively the corresponding $P$ using Eq.\ (\ref{eq:tp}). We measure:

\begin{tabular}{c|c|c|c|c|c|c|c|c}
P(1) & P(2) & P(3) & T(2;1) & T(3;1) & T(1;2) & T(3;2) & T(1;3) & T(2;3)\\
\hline \hline 3/5 & 1/5 & 1/5 & 1/8 & 2/8 & 1/8 & 1/8 & 2/8 & 1/8
\end{tabular}
\\[2mm]
Notice that we counted for each type of path both directions separately. Solving Eq.\ (\ref{eq:tp}) for $P(k_1|k_2)$ then yields:

\begin{tabular}{c|c|c|c|c|c}
P(2$|$1) & P(3$|$1) & P(1$|$2) & P(3$|$2) & P(1$|$3) & P(2$|$3)\\
\hline \hline 1/3 & 2/3 & 1/2 & 1/2 & 2/3 & 1/3
\end{tabular}
\\[2mm]
In the same way we can proceed to measure elements $T(k_3;k_2;k_1)$ etc.

\paragraph*{Generalized Maslov-Sneppen}
In \cite{43,44} Maslov and Sneppen introduce a local rewiring algorithm, that randomizes correlated networks, while strictly preserving the degrees of the nodes. It yields networks with an identical topology as MR-networks. We propose a generalization of this MS-algorithm, that randomizes networks
while preserving not only the degree distribution but also correlations up to a certain range. This algorithm should be especially useful in the examination, which correlations are important and which correlations can be neglected in real networks. For example, it will be helpful to answer a question of the following type: How far is the range of correlations that must be considered to determine the percolation threshold of a certain network up to a certain error?
\begin{figure}[tb]
\includegraphics[width=0.47\textwidth]{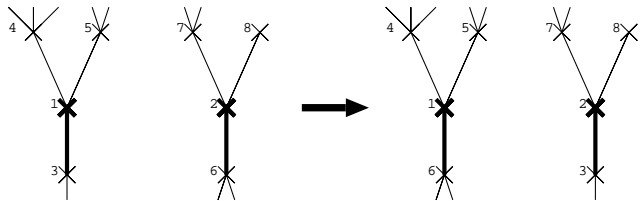}
\caption{Generalized MS-algorithm: randomization conserving correlations of range up to one.}
\label{fig:genms2}
\end{figure}
Fig.\ \ref{fig:genms2} presents the algorithm where the network is randomized regarding all correlations of range longer than one, i.e.\ only the correlations between next neighbors are kept:
\begin{itemize}
\item Choose randomly one motif link-node (with node-degree $k$) in the network.
\item Choose another motif that has the same topology link-node (with the same degree $k$).
\item Exchange the parts attached to the end of the link. 
\end{itemize}
A motif consists of a small connected neighborhood and possibly dangling edges attaching to some nodes of the neighborhood. We see that in the process the distribution of paths of length one stays the same. However, the distributions of longer paths are randomized.

\begin{figure}[tb]
\centering
\includegraphics[width=0.47\textwidth]{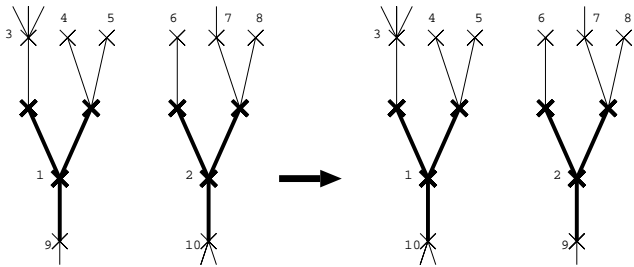}
\caption{Generalized MS-algorithm: randomization conserving correlations of range up to two.}
\label{fig:genms3}
\end{figure}

This can easily be generalized to randomization where correlations of range up to $x$ are preserved. Just exchange in the prescription above the motif link-node with a tree link-node-...-link-node of range $x$ (i.e. each branch has $x$ links and nodes). The rest of the process stays the same. For example, when the motif consists of a tree with range two, then the distribution of paths of length two stays the same. With respect to further-ranging correlations the network is again randomized (cp. Fig. \ref{fig:genms3}).
Finally, the known MS-algorithm is obtained, when the motif is only a link. Then only the degree distribution of the network and none of the correlations are preserved.
 
\begin{figure}[tb]
\centering
\includegraphics[width=0.47\textwidth]{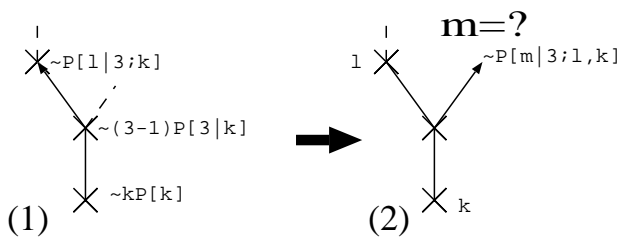}
\caption{Constructing a network with correlations of range two.}
\label{fig:build}
\end{figure}

\paragraph*{Creating networks with next-neighbor correlations}
It is fairly easy to create a network with correlations only between next neighbors. Such a network is determined by the degree distribution and all correlations $P(k|l)$, where $k>l$. All other $P(k|l)$ are fixed either by the normalization condition $\sum_k P(k|l)=1$ or by (\ref{paths}), which amounts to: $P(l) l P(k|l)=P(k) k P(l|k)$. Now, the network can be created in the following way:
\begin{itemize}
\item We give $N$ nodes distributed according to $P(k)$.
\item We randomly connect the stubs of all nodes with degree 1 with stubs of other nodes according to $P(k|1)$.
\item We link all nodes with degree 2 according to $P(k|2)$ (with $k \geq 2$ of course) etc.
\end{itemize}
According to this procedure it is much more difficult to construct networks with correlations ranging further. We want to shortly illustrate the difficulties arising from the generalization by examining the case of correlations of range two. In addition, we allow only nodes with degrees up to three in the network. In Fig.\ \ref{fig:build} we show how such a
network is constructed. Step (1) in the figure is unproblematic, where starting from node $k$ we build a chain using $P(3|k)$ and $P(l|3;k)$ (cf. Eq.\ (\ref{eq:tp})). For step (2) however we have to respect that the degree $m$ is determined by both of the node-degrees $k$ and $l$. So, we have to introduce a new correlation function to determine $m$: $P(m|3;k,l)$. Similar to Eq.\ (\ref{paths}), again several consistency equations need to be fulfilled:
\begin{align}
&2NP(k) k P(3|k) P(l|3;k) P(m|3;l,k)=& \nonumber \\
=&2NP(k) k P(3|k) P(m|3;k) P(l|3;m,k)=& \nonumber \\
=&2NP(m) m P(3|m) P(k|3;m) P(l|3;k,m)=& \nonumber \\
=&2NP(m) m P(3|m) P(l|3;m) P(k|3;l,m)=& \nonumber \\
=&2NP(l) l P(3|l) P(m|3;l) P(k|3;m,l)=& \nonumber \\
=&2NP(l) l P(3|l) P(k|3;l) P(m|3;k,l).& 
\label{eq:trihedral}
\end{align}
Each term calculates the number of trihedrals in the network with a node of degree 3 as center and $k$, $l$, $m$ as the degrees of the other nodes. Of course this number must not depend on the order in which we develop the nodes of the trihedral. The probability distribution for the trihedral motifs is denoted $T(3;k,l,m)$. Eq.\ (\ref{eq:trihedral}) then corresponds to the fact that all permutations of $k$, $l$, $m$ yield the same $T(3;k,l,m)$. We get the following set of relations:
\begin{eqnarray*}
P(k|3;l,m)&=&P(k|3;m,l)
\label{eq:cond3fst}\\
P(l|3;k) P(m|3;l,k) &=& P(m|3;k) P(l|3;m,k).
\label{eq:cond3snd}
\end{eqnarray*}
Due to the symmetry of the problem, it is not surprising that we need only one of the correlation functions $P(k|3;l,m)$. Now, we have all conditional probabilities necessary and can construct a network with correlations of range two and maximum degree three:
\begin{itemize}
\item Construct all chains $T(k;2;l)$ according to Eq.\ (\ref{eq:tp}). Divide the amount of all resulting motifs by $2 \times 1$ (number the same motif will be developed by the correlation functions).
\item Construct all trihedrals $T(3;m,l,k)$. Starting with Eq.\ (\ref{eq:tp}) we find all $T(k;3;l)$, then we continue with $T(3;m,l,k)=P(m|3;l,k)T(l;3;k)$. Divide the resulting number of all motifs by $3 \times 2 \times 1$. 
\item Combine all motifs randomly, e.g.\ a trihedral $T(3;2,l_2,k_2)$ and a chain $T(k_1;2;3)$ or two chains $T(k_1;2;2)$ and $T(2;2;k_2)$ or two trihedrals $T(3;m_1,3,k_1)$ and $T(3;3,l_2,k_2)$.
\end{itemize}
A generalization of this concept is straight-forward, but gets complex quite fast.

\section{Percolation condition}
\label{perer}
\subsection{Theoretical derivation}
\begin{figure}[tb]
\centering
\includegraphics[width=0.47\textwidth]{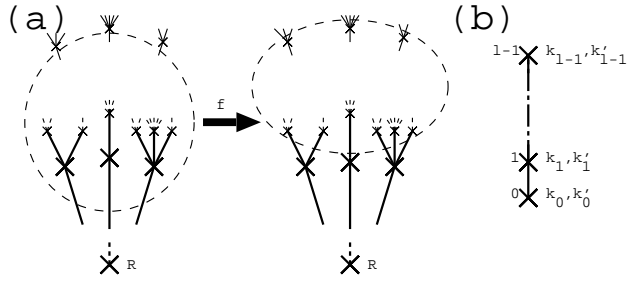}
\caption{(a) Graphical representation of the map Eq.\ (\ref{abb:f}): The dotted circle moves away exactly one step from an arbitrarily chosen root of the giant cluster. The dotted circle on the left side circumscribes a sufficiently large connected part of the giant component. Now, only at the percolation transition, $f$ reproduces on the right side exactly the same topology, including the same number of motifs and the same distribution. (b) An exemplary motif, a chain with degree sequence $k_0$, $k_1$, ..., $k_{l-1}$. A motif belongs to the dotted circle in (a) if its first node 0 is in it.}
\label{fig:abb}
\end{figure}

In \cite{cohetal, cohetal2} Cohen {\it et al.} introduce as a percolation condition for the MR-model:
\begin{center}
\parbox{8cm}{A graph has a spanning cluster, when a site j, which is reached by following a link from site i on the giant cluster, has at least one other link on average.} 
\end{center}
To be applicable to our correlated random networks we generalize this condition: 
\begin{center}
\parbox{8cm}{A correlated graph is characterized by the distribution of motifs, which fully determines the topology of the graph. A giant component exists, if motifs which are linked to the giant cluster have at least one other link (leaving the motif) on average.} 
\end{center}
In our model with correlations of range $l-1$ the network is characterized by the distribution of trees with branches of length $l-1$ (Fig.\ \ref{fig:genms2}: node-link-node and Fig.\ \ref{fig:genms3}: node-link-node-link-node). These trees are the motifs of our generalized random model with correlations of range up to $l-1$. For the calculation of the percolation threshold $p_c$ we do not consider the whole trees as motifs but confine to chains of length $l-1$ (i.e. $l$ nodes and $l-1$ links, cp.\ Fig.\ \ref{fig:abb}b). All solutions obtained for $p_c$ are also solutions for the whole trees as motifs instead of the chains. We define the map
\begin{eqnarray}
&& f\left [T_{GC,pc}^*(k_{l-1},k_{l-1}';...;k_1,k_1';k_0,k_0') \right]:=\sum\limits_{k_0,k_0'}(k_{l-1}'-1){k_{l}-1 \choose k_{l}'-1}p_c^{k_{l}-k_{l}'}\times 
\nonumber \\  && \qquad
\times (1-p_c)^{k_{l}'-1}T_{GC,pc}^*(k_{l-1},k_{l-1}';...;k_1,k_1';k_0,k_0')P(k_{l}|k_{l-1};...;k_1). \label{abb:f}
\end{eqnarray}
Here, $T_{GC,pc}^*$ is the normalized probability distribution of chains in the giant component (cp.\ Fig.\ \ref{fig:abb}b) with degree sequence $k_0,...,k_{l-1}$ in the original network and degree sequence $k_0',...,k_{l-1}'$ in the percolated network. $P(k_l|k_{l-1};...;k_1)$ determines the degree $k_l$ of node $l$ that is next in the chain. Again, $k_l$ is the original degree, and ${k_{l}-1 \choose k_{l}'-1}(1-p_c)^{k_{l}'-1} p_c^{k_{l}-k_{l}'}$ determines the probability that the percolated degree is $k_l'$. At last, the factor $k_{l-1}'-1$ accounts for the number of chains developed due to the number of outgoing links of node $l-1$. In Fig.\ \ref{fig:abb}a we see a graphical representation of this map (\ref{abb:f}). Due to the percolation condition introduced at the beginning of this section, at the threshold $p_c$ the following relations
must be fulfilled:
\begin{enumerate}
\item On the right side there must be the same distribution of motifs in the dotted circle as on the left side.
\item On the right side there must be the same number of motifs in the dotted circle as on the left side.
\end{enumerate}
These conditions translate into the following set of equations: 
\begin{equation}
f\big[T_{GC,pc}^*(k_{l-1},k_{l-1}';...;k_1,k_1';k_0,k_0')\big] 
=T_{GC,pc}^*(k_{l},k_l';...;k_2,k_2';k_1,k_1'),\label{final5}
\end{equation}
\begin{equation}
\sum\limits_{k_0,k_0',...,k_{l-1},k_{l-1}'} (k_{l-1}'-1) T_{GC,pc}^*(k_{l-1},k_{l-1}';...;k_0,k_0')=1.
\label{final6}
\end{equation}
Notice, that in the first equation the distribution on the right side is only normalized in case that Eq.\ (\ref{final6}) is fulfilled.
Eq.\ (\ref{final6}) states that the number of motifs, which a motif connects to, is one in average. A short derivation yields the following set of equations:
\begin{align}
\sum\limits_{k_1} (1-p_c)(k_{l}-1)T_{GC,pc}(k_{l-1};...;k_2;k_1)
P(k_{l}|k_{l-1};...;k_1)&=T_{GC,pc}(k_{l};...;k_3;k_2),
\label{final1}\\
\sum\limits_{k_1,...,k_{l-1}} T_{GC,pc}(k_{l-1};...;k_2;k_1)&=1.
\label{final2}
\end{align}
Notice, that the $T$ are different from the $T^*$ in Eq.s (\ref{final5}) and (\ref{final6}). Here, we interpret the $T$ only as auxiliary variables (the definition of it can be seen when comparing Eq.s (\ref{final6}) and (\ref{final2})). Eq.s (\ref{final1}) and (\ref{final2}) are a system of $K^{l-1}+1$ equations, where $K$ is the number of different degrees in the considered network and $l-1$ is the range of the correlations. There are $K^{l-1}+1$ variables: $p_c$ and $T_{GC,pc}(k_{l-1};...;k_1)$, where $k_1$,...,$k_{l-1}$ take on the values of all $K$ degrees. Thus, the set of equations (\ref{final1}) and (\ref{final2}) allows to determine $p_c$.

\paragraph*{Node or link removal} 
Considering these generalized random networks in the thermodynamic limit, the percolation threshold is the same for node and edge percolation. Edge percolation removes the fraction $p$ of all links from the network. Node percolation removes $p$ of all nodes from the network, i.e. the network size is changed from $N$ nodes to $(1-p)N$ nodes, and from the remaining $(1-p)N$ nodes those $p$ links are removed, which connect to removed nodes. We can thus formulate the relation between the percolated degree distribution under link removal $P_{lr}(k',k)$ and that under node removal $P_{nr}(k',k)$:
\begin{equation} 
P_{nr}(k',k)=(1-p)P_{lr}(k',k)+p\;\delta_{k',0}P(k).
\label{rel}
\end{equation}
Here, $k$ is the original degree and $k'$ the percolated. The probability, that the link between node x-1 and node x is removed in edge percolation, is $p$. In node percolation we have the same probability $p$, that the node x is removed and consequently also the link connecting node x-1 with x. All conditional probabilities are the same, e.g. 
\begin{equation} 
P_{nr}(k_x|k_{x-1};...;k_1)=P_{lr}(k_x|k_{x-1};...;k_1). \label{rel2}
\end{equation}
These are all original degrees. We see, that the distribution of motifs is the same at every $p$ for both networks with different percolation procedures. Only the network size differs by a factor $1-p$ due to (\ref{rel}). Hence, both procedures yield the same percolation threshold.

\subsection{Examples}
\paragraph*{Exact ones}
We have tested the Eq.s (\ref{final1}) and (\ref{final2}) with different networks using \textsc{Mathematica}$^{\text{\tiny\textregistered}}$:
\begin{enumerate}
\item $P(k)=\frac{1}{3} \delta_{k,1}+\frac{1}{3} \delta_{k,2}+\frac{1}{3} \delta_{k,3}$ and 
\\$P(k|l)=\delta_{l,1} \delta_{k,3}+\delta_{l,2}(\frac{1}{2} \delta_{k,2}
+ \frac{1}{2} \delta_{k,3})+ \delta_{l,3}(\frac{1}{3} \delta_{k,1}+\frac{1}{3} \delta_{k,2}+\frac{1}{3} \delta_{k,3})$
\item $P(k)=\frac{1}{3} \delta_{k,1}+\frac{1}{3} \delta_{k,2}+\frac{1}{3} \delta_{k,3}$ and\\
 $P(k|l)=\delta_{l,1} \delta_{k,2}+\delta_{l,2}(\frac{1}{2} \delta_{k,1}+\frac{1}{2} \delta_{k,3})+
\delta_{l,3}(\frac{1}{3} \delta_{k,2}+ \frac{2}{3} \delta_{k,3})$ and\\
$P(k|l;m)=\delta_{m,1} \delta_{l,2} \delta_{k,3}+ \delta_{m,2} \delta_{l,3} 
(\frac{1}{3} \delta_{k,2}+\frac{2}{3} \delta_{k,3})+ \delta_{m,3} \delta_{l,2} \delta_{k,1}+ \delta_{m,3} \delta_{l,3} (\frac{1}{3} \delta_{k,2}+ \frac{2}{3} \delta_{k,3})$
\end{enumerate}
We find for the networks:
\begin{enumerate}
\item $p_c=\frac{1}{7}$
\item $p_c=\frac{1}{4}$
\end{enumerate}
To check if our algorithm yields the correct results, we use the correspondence of these networks to MR-networks:
\begin{enumerate}
\item $P(k)=\frac{1}{2} \left ( \frac{2}{3} \right ) ^3 \delta_{k,3} +\left [ \frac{1}{2}+\frac{3}{2} 
\frac{1}{3}\left ( \frac{2}{3} \right ) ^2 \right ] \delta_{k,2}+\frac{3}{2} 
\frac{2}{3}\left ( \frac{1}{3} \right ) ^2 \delta_{k,1}+\frac{1}{2} \left ( \frac{1}{3} \right )^3 \delta_{k,0}$
\item $P(k)=\left ( \frac{2}{3} \right ) ^3 \delta_{k,3} +\left [ 3 
\frac{1}{3}\left ( \frac{2}{3} \right ) ^2 \right ] \delta_{k,2}+3 
\frac{2}{3}\left ( \frac{1}{3} \right ) ^2 \delta_{k,1}+\left ( \frac{1}{3} \right )^3 \delta_{k,0}$
\end{enumerate}
The first example exhibits no correlations except that nodes with degree $k=1$ connect only to those with degree $k=3$. This means that with probability $1/3[P(1)/P(3)]=1/3$ a stub of a node with degree 3 is blocked. We can replace each node with $s$ blocked stubs by a node with degree $k=3-s$. In this `new' network there are no correlations at all. For this MR-network the percolation threshold is calculated according to \cite{cohetal,mollreed}
\begin{equation}
\label{eq:percthrs}
1-p_c=\frac{1}{\frac{\lgl k_0^2 \rgl}{\lgl k_0 \rgl}-1}.
\end{equation}
The result $p_c=1/7$ corresponds. In the second example again stubs are blocked, but this time by chains consisting of one node with degree $k=2$ and one with degree $k=1$. Again, there are no other correlations. We find the correct $p_c=1/4$. 

\paragraph*{MR-model}
Our method yields the threshold of the MR-network correctly. There, the trivial correlations are
\begin{equation}
P(k_l|k_{l-1};...;k_1)=\frac{k_l P(k_l)}{\langle k\rangle}.
\label{corr}
\end{equation}
Employing this in (\ref{final1}), summing over $k_2$,..., and $k_l$ we get with (\ref{final2})
\begin{eqnarray}
&& \sum_{k_l} (1-p_c)(k_l-1)\frac{k_l P(k_l)}{\langle k\rangle}
\stackrel{!}{=}1. \label{proof2}
\end{eqnarray}
This is identical with (\ref{eq:percthrs}), the correct percolation threshold for uncorrelated random networks.

\begin{figure}[tb]
\centering
\includegraphics[width=0.47\textwidth]{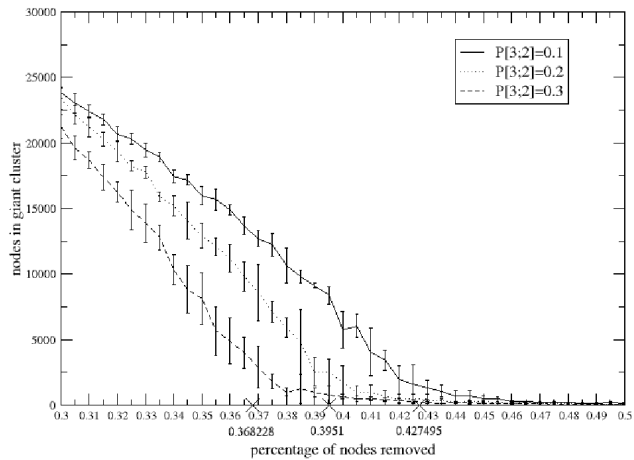}
\caption{Percolation in a random network with pair correlations. 
The network size was 100,000 nodes. For all networks we gave the degree distribution as $P(1)=P(2)=P(3)=1/3$ and the correlations $P(2|1)=0.3$, $P(3|1)=0.2$. The different curves represent different correlations $P(3|2)$, which takes on the values 0.1, 0.2, and 0.3. All points are averaged over six runs. The crosses on the x-axis are the theoretical values for the percolation thresholds calculated according to Eq.s (\ref{final1}) and (\ref{final2}).}
\label{fig:simgenran}
\end{figure}

\paragraph*{Simulations}
We also tested the algorithm numerically using correlations that cannot be mapped on an uncorrelated network (view Fig.\ \ref{fig:simgenran}). We measured the size of the giant component in percolated networks with different correlation profiles. The size of the networks we created was 100,000 nodes. The degree distribution of the original network at $p=0$ was always the same $P(1)=P(2)=P(3)=1/3$, also the neighbor correlations $P(2|1)=0.3$ and $P(3|1)=0.2$. For $P(3|2)$ we employed the values 0.1, 0.2, and 0.3. As noted before, all other correlations are determined by these. We calculated for the percolation transition the values: 0.368, 0.395, and 0.427 respectively. With the completely different method of V\'azquez and Moreno \cite{Vaz} the same results were obtained. In Fig.\ \ref{fig:simgenran} these values correspond quite well with the simulations. Of course, it is impossible to determine exactly the $p_c$ from the simulations.

\paragraph*{Note}
We have seen in the exact examples above 1.) and 2.) that correlations play a role for the percolation properties of networks, since both networks have the same degree distribution, but different correlations and different percolation thresholds. For the uncorrelated network with the same degree distribution we get according to (\ref{eq:percthrs}): $p_c=1/4$. Example 1.) yields a smaller $p_c=1/7$ due to the dissortativity of the network: nodes with degree $k=1$ only link to nodes with degree $k=3$. An example for assortative pair correlations would be: $P(k|l)=\delta_{l,3}\delta_{k,3}+\delta_{l,2}\delta_{k,2}+\delta_{l,1}\delta_{k,1}$. There are two giant components in such a network and thus two different thresholds $p_{c,1}=0$ and $p_{c,2}=\frac{1}{2}$. Thus, there exists a giant component for all $p<1/2$, which is then by definition the value of the percolation threshold for the whole network. That means that assortativity yields a higher $p_c$ in comparison to the uncorrelated network.

\section{Outlook: loops and clustering}

\begin{figure}[tb]
\centering
\includegraphics[width=0.47\textwidth]{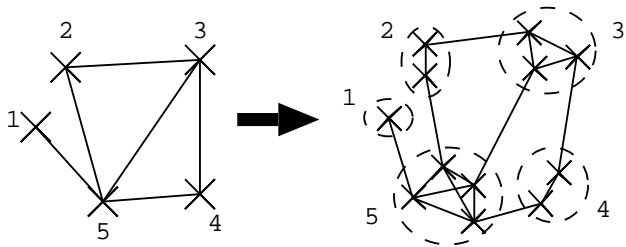}
\caption{An example for an algorithm, that decreases the power-law exponent $\gamma$ of the degree distribution by one without increasing the percolation threshold: All nodes i in the network with degree $k_i$ are replaced by neighborhoods of $k_i$ nodes with degree $k_i$. For each node $k_i-1$ of its links connect to the other $k_i-1$ nodes in the neighborhood and one link connects to the outside of the neighborhood.}
\label{fig:NBA}
\end{figure}
In this section, we show that the percolation properties depend not only on the degree distribution and the degree correlations, but also on the length of typical loops in the network. Note that loops cannot be sufficiently described by degree correlations (for our generalized correlated random model, the appearence of certain loops depends crucially on the network size). This problem is also studied in the context of embedding a network into a geography \cite{geog,Soki,ramon}.

We consider the following example: a MR-network with a power-law degree distribution $P_i(k)$ with an exponent $\gamma >3$. This network has a percolation threshold $p_{c,i}<1$ (cf.\ Eq.\ \ref{eq:percthrs}). Replacing links by motifs link-node-link, we integrate new nodes in the network in a manner that the resulting degree distribution $P_f(k)$ follows a power-law with an exponent $\g \leq 3$. We consider these two possibilities:
\begin{enumerate}
\item We integrate the new nodes in a way, that the resulting network corresponds to a MR-model.
\item We assign a certain number $N_i$ of the new nodes to each existing node i with degree $k_i$. With those $N_i$ nodes we build a neighborhood around i that has exactly $k_i$ outgoing links (cf.\ for example Fig.\ \ref{fig:NBA}).
\end{enumerate}
This yields the following percolation thresholds:
\begin{enumerate}
\item $p_c=1$, because the second moment of the degree distribution $P_f(k)$ diverges. 
\item $p_c<1$, because the resulting network with degree distribution $P_f(k)$ has a superstructure: a network with degree distribution $P_i(k)$ where the nodes are finite-size neighborhoods. This superstructure determines an upper boundary for the percolation threshold.
\end{enumerate} 
Note that it might well be possible to ensure that both networks also have the same degree correlations. So, apparently it plays an important role for the value of the percolation threshold, if additional links are only added to a neighborhood or connect distant parts of the network. Consequently, it is an interesting question, how many links of the Internet or social networks connect only nodes that already belong to the same neighborhood. We stress that contrary to the BA- and MR-models real networks seldom exhibit local tree-structure, indicating redundant linking in neighborhoods. This points to common interests or properties of nodes. For example in social networks these may be: geographic location, language, age or level of education.

\subsection{Extension of the BA-model}
It is an obvious failure of the BA-model, that it does not implement clustering. The clustering coefficient tends to zero in the thermodynamic limit (number of triangles divided by number of pairs of adjacent edges, i.e.\ edges with at least one identical endpoint). We suggest here a simple extension of the BA-model that allows to influence clustering, while it preserves the degree distribution of the BA-model.

We assume that every new node added to the network brings with it $m=2$ proper links. These proper links connect the new node with nodes in the network according to different criteria. For example, in a friendship-network, every individual would have the right to choose two friends. The first friend he chooses from people who do the same job as he. The second friend he chooses from people who have the same favorite hobby. Both times he preferably chooses those people that already have a lot of friends (i.e.\ preferential attachment).

The new feature compared with the BA-model is, that at its introduction we assign to each node $i$ two parameters, a job-parameter $0 \leq p_{j,i} \leq 1$ and a hobby-parameter $0 \leq p_{h,i} \leq 1$. Each new node has a job-link and a hobby-link. Now, according to preferential attachment we first determine the degree $k_1$ of node 1, that the job-link shall attach to. Then we search for that node 1 that has the $p_{j,1}$ closest to $p_{j,i}$, corresponding to the best matching of common interests. The same procedure determines node 2, that the hobby-link of $i$ attaches to.

Qualitatively, the clustering depends on the correlations between the parameters $p_{j,i}$ and $p_{h,i}$. We choose $p_{j,i}$ uniformly at random. $p_{h,i}$ is chosen depending on the value $p_{j,i}$. There are two limiting cases:
\begin{enumerate}
\item The choice of $p_{h,i}$ is independent from the choice of $p_{j,i}$. Then our model corresponds exactly to the BA-model with $m=2$ and exhibits a vanishing clustering coefficient.
\item $p_{h,i}=p_{j,i}=p_i$. Then the clustering is maximal.
\end{enumerate}
For the second case, when the degrees $k_1$ and $k_2$ of nodes 1 and 2 are equal a double bond is formed. If as an additional rule we prohibit double bonds, the second link shall be connected to a node 2 with the parameter closest to $p_i$ but unequal node 1. In this model the clustering is maximal, because the probability that nodes 1 and 2 are neighbors is maximal. In the case that 1 and 2 are neighbors a new triangle is formed in the network. 

Another important feature of the second limiting case is that the clustering coefficient is independent of the network size. This is proven in the following way: The probability, that a node $x$ of degree $k_x$ has a neighbor of degree $k_y$, depends only on the degrees of the nodes and not on the network size. If node $x$ has a neighbor $y$ of degree $k_y$, then the distribution for the difference in parameters $P(|p_x-p_y|/N)=P(|p_{h,x}-p_{h,y}|/N)=P(|p_{j,x}-p_{j,y}|/N)$ is independent of the network-size. This guarantees that the probability, that the two nodes 1 and 2 (in the growth algorithm presented above) are connected, is independent of the network size. Thus, the probability, that with a new node also a new triangle is added to the network is independent of the network size. From the fact, that with every node three pairs of adjacent edges are added to the network and with a constant probability sometimes a triangle, it follows, that the clustering coefficient is independent of $N$ (at least for $N$ large).

A generalization of this model is straightforward. Let $P(k_x,k_y,|p_{j,x}-p_{h,y}|/N)$ be the probability, that two nodes $x$ and $y$ with degrees $k_x$ and $k_y$ and with difference in parameters $|p_{j,x}-p_{h,y}|$ are neighbors. If the parameter $p_{h,i}$ for the new nodes is chosen such, that $P(k_x,k_y,|p_{j,x}-p_{h,y}|/N)$ stays constant with growing network size $N$, then the clustering coefficient is approximately independent of $N$. This should be the case for a scale-invariant probability distribution for the choice of $p_{h,i}$ for node i: $P(p_{h,i})=f \left ( |p_{h,i}-p_{j,i}|/N \right)$ with some arbitrary function $f$. Let us finally note, that this is a plausible condition for friendship networks, but also for many other networks. The pool of possible friends, from which a new person can choose, is always independent of the network size. When somebody comes to a new city to make friends, for his choice, it does not matter, how many people are living on the whole earth. He will choose his friends from his immediate neighborhood (mathematically these neighborhoods are characterized by similarity in the parameters $p$). The only difference is that with growing network size, i.e.\ population on the earth, the parameter interval is renormalized. The parameters of people in one city then lie closer together, the differences are simply scaled. Empirically, the network size $N$ does not influence the person's choice of friends, though.  

The parameters $p_j$ and $p_h$ can of course be identified with two-dimensional coordinates in a geography. A generalization to more than two parameters is straight-forward. In summary we can say that the clustering in our example network is larger, when people with the same job tend to have the same hobbies. This very general outline was supposed to show the origin of network-size-invariant clustering and appearance of neighborhoods (with loops) in real networks.

\section{Conclusion}
We discussed a general model for correlated random networks -- where correlations can have arbitrary range. We presented two algorithms to produce or examine those networks. One, a generalization of a randomization algorithm by Maslov and Sneppen, should be helpful in examining the influence of correlations in real networks. Especially the question, which range of correlations is important, can be answered. We derived a set of equations  (\ref{final1}) and (\ref{final2}) determining the percolation threshold for this model and verified the result by different methods. In the end we added a few general remarks concerning clustering -- a measure of the amount of triangles in a network -- and more generally concerning the existence of loops. For both correlations and loops their influence on topological properties like the percolation threshold of a network is not yet fully understood. We suggested very shortly a mechanism for the emergence of network-size-invariant clustering in real networks.

\end{document}